# Photonic microwave generation with high-power photodiodes

Tara M. Fortier[1], Franklyn Quinlan[1], Archita Hati[1], Craig Nelson[1], Jennifer A. Taylor[1], Yang Fu[2], Joe Campbell[2], Scott A. Diddams[1]

[1] National Institute of Standards and Technology, Time and Frequency Division, 325 Broadway MS 847, Boulder CO 80305
[2] Department of Electrical and Computer Engineering, University of Virginia, Charlottesville, VA 22904 USA
*Corresponding author: fortier@boulder.nist.gov, sdiddams@boulder.nist.gov

We utilize and characterize high-power, high-linearity modified uni-traveling carrier (MUTC) photodiodes for low-phase-noise photonic microwave generation based on optical frequency division. When illuminated with picosecond pulses from a repetition-rate-multiplied gigahertz Ti:sapphire modelocked laser, the photodiodes can achieve 10 GHz signal power +14 dBm. Using these diodes, a 10 GHz microwave tone is generated with less than 500 attoseconds absolute integrated timing jitter (1 Hz – 10 MHz) and a phase noise floor of –177 dBc/Hz. We also characterize the electrical response, amplitude-to-phase conversion, saturation and residual noise of the MUTC photodiodes.

A photonics approach to low noise microwave generation has enabled 10 GHz signals with close-to-carrier absolute phase noise less than -100 dBc/Hz, 40 dB lower than the best room-temperature electronic oscillators [1]. The exceptionally low phase noise arises from the frequency division of a stable optical oscillator [2, 3] using a femtosecond laser frequency comb [1, 4, 5]. This new development in low-noise microwave generation is anticipated to impact a variety of fields, including basic spectroscopy [6], radar and sensing [7], and high-speed analog-to-digital conversion [8].

In optical frequency division (OFD), the stability of the optical reference is transferred to the timing in the pulse train of a modelocked laser. Photodetection of this stable optical pulse train generates an electronic pulse train with a spectrum composed of high spectral purity harmonics of the pulse repetition rate, up to the cut-off bandwidth of the photodiodes (PD). Unfortunately, photodiode nonlinearities can degrade spectral purity by coupling amplitude noise on the optical intensity to phase noise on the microwave signal (AM-PM), as well as restricting the maximum signal strength. In this Letter, we directly address these two issues in photodetection with the use of modified uni-traveling carrier (MUTC) photodiodes for low-noise microwave generation. Under short pulse illumination, the high-power MUTC photodiodes provide 10 GHz carrier strengths up to +14 dBm. This represents greater than 30× increase over previous results with other photodiodes [1, 9, 10], and facilitates a phase noise floor of –177 dBc/Hz. At the same time, the MUTC photodiodes exhibit high linearity, leading to a 10 GHz AM-PM coefficient below 0.1 rad for photocurrents up to 20 mA. This limits the impact of laser relative intensity noise (RIN), which is key to our achievement of absolute 10 GHz integrated timing jitter of 500 attoseconds (1 Hz to 10 MHz). Additionally, the residual noise of the MUTC photodiodes exhibits a close-to-carrier flicker noise below -120 f$^{-1}$ dBc/Hz.

The impact of PD saturation and AM-PM on microwave generation is of particular concern under pulsed illumination because peak optical powers can be greater than 1000 times higher than that under continuous wave illumination [11-14]. This creates high densities of photo-generated carriers in the PD absorption layer which, along with finite carrier mobility in the depletion layer, produces a space charge with an electric potential counter to the applied reverse bias. This intensity-dependent charge screening acts to slow down the response time of the PD and induces AM-PM. Optical filtering and pulse interleaving have been shown to increase the photodiode linearity [9, 15, 16], allowing for extraction of larger microwave signals while simultaneously enabling lower AM-PM conversion. To further improve microwave power and linearity, here we utilize the MUTC photodiode structure to effectively suppresses the space charge effect [17]. The MUTC PDs employ an InP collector that is doped to predistort the static internal electric field; this prevents field collapse at the heterojunction interface at high current. A portion of the absorber is undepleted in order to maintain high field across the heterojunctions. The structure also has a charge or "cliff" layer that tailors the electric field profile to prevent the field in the depleted absorber from collapsing at high photocurrent. MUTC PDs have produced microwave power at 15 GHz approaching 1 W when illuminated with amplitude modulated CW light at 1550 nm [18]. Here, we explore the MUTC PD response under picosecond pulse illumination at 980 nm, focusing on the phase noise of the generated microwaves.

Figure 1(a) shows the 10 GHz carrier strength as a function of generated photocurrent for a 56 μm diameter MUTC PD when operated at low (9 V) and high (21 V) bias voltage. The MUTC PD was anti-reflection coated and exhibited a responsivity of ~0.3 A/W when illuminated in free-space with chirped 1 ps pulses at 980 nm from a 1 GHz Ti:sapphire laser [19]. A microwave probe and a bias-T are used for applying a reverse bias and extracting the generated microwave signal. Both the 10 GHz microwave power and AM-PM conversion coefficient were measured under illumination by 1 and 2 GHz pulse trains. Repetition rate doubling is achieved with a free-space Mach-Zehnder pulse interleaver that is similar to previous fiber-optic implementations [15, 16].

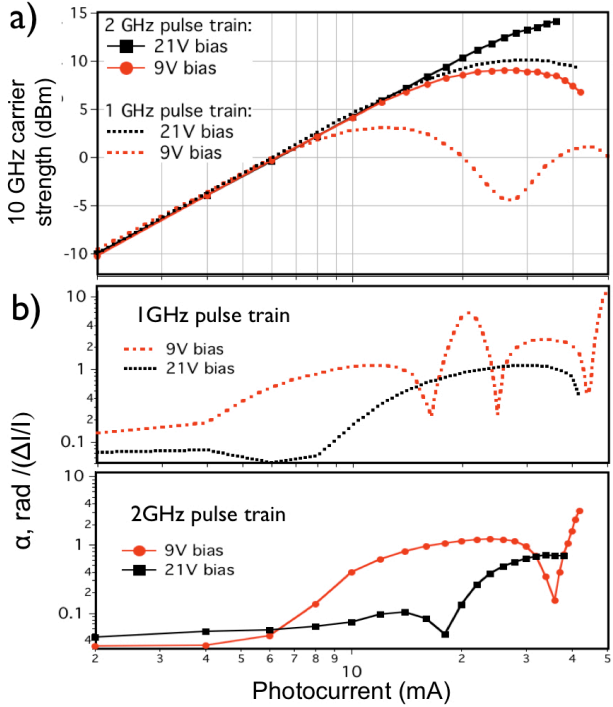

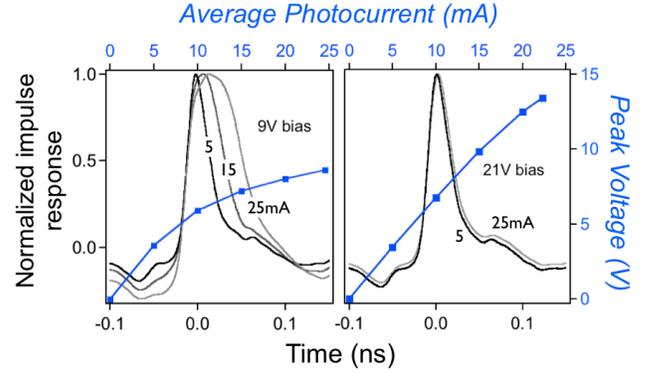

Fig. 1. (Color online) Saturation and AM-PM measured for a 10 GHz carrier obtained from MUTC PD with 1 ps pulses at 980 nm from a modelocked laser. $\alpha$ is the rms 10 GHz phase variation per fractional change in the photocurrent ($\Delta I/I$).

With 21 V bias and approximately 125 mW of the 2 GHz pulse train incident, the MUTC PD generated a 10 GHz carrier with +14 dBm at 37 mA of average photocurrent. With the 1 GHz pulse train, up to 50 mA of photocurrent was generated with approximately 170 mW of average laser power. At these powers, however, the PD was strongly saturated. In Fig. 1(b), we present the AM-PM coefficient, $\alpha$, which is a conversion factor between phase noise on the 10 GHz carrier and RIN on the optical pulse train [12]. We measured $\alpha$ for the MUTC PD in a manner similar to that in Ref. [14]. We observe that improved linearity at 2 GHz and higher bias voltage result in a lower overall value of $\alpha$. In fact, only at photocurrents when the microwave power begins to saturate does $\alpha$ deviate significantly from zero. This can be contrasted to our previous measurements of AM-PM with lower power PDs, where we observed both larger values and variations of $\alpha$ for photocurrents < 8 mA. While the presence of nulls in $\alpha$ can serve as optimal operating points for low noise microwave generation, their positions are sensitive to operating conditions and can change slowly with time [12]. On the other hand, with the MUTC PDs operated at 21 V and a 2 GHz pulse train, $\alpha$ is < 0.1 rad for all photocurrents less than 20 mA. With the AM-PM contribution to the single sideband phase noise being, $L_{RIN}(f) = RIN + 20 \log (\alpha) - 3\ dB$, a value of $\alpha$ < 0.1 rad would result in phase noise 23 dB lower than the RIN on the optical pulse train.

The improved linearity at high bias voltage is also seen in the electrical response in the time domain. As shown in Fig. 2, high bias voltage helped to reduce broadening and temporal shifts of the electronic pulses at high photocurrent. This reduced distortion pushes both saturation of the 10 GHz carrier and large AM-PM to higher incident average optical power.

Fig. 2. (Color online) Normalized electrical response and the peak voltages obtained with MUTC photodiode for different average photocurrents when illuminated with a 980 nm, 2 GHz pulse train.

The high signal strengths at 10 GHz and low AM-PM conversion indicate that MUTC PDs would be excellent detectors in photonic microwave generators. As a test of their performance, we use MUTC PDs to generate a microwave signal from a stable pulse train. Figure 3 shows both the residual noise added in photodetection, as well as the absolute noise of the 10 GHz signal. The residual measurement was obtained by splitting the pulse train from a single OFD, after the Mach-Zehnder interleaver, and illuminating two separate photodiodes. The generated 10 GHz signals were then compared using cross-correlation spectral analysis to determine the phase noise [20, 21]. The residual measurement characterized the phase noise added in photodetection under the operating conditions of a 21 V applied bias and 18 mA photocurrent. These conditions yielded a 10 GHz power of +10 dBm. Although insensitive to common mode noise on the OFD, this measurement reflects noise contributed by AM-PM conversion, shot noise, flicker, and thermal noise.

The most important features in the residual phase noise measurement are flicker noise of approximately -122 $f^{-1}$ dBc/Hz, and the high-frequency noise floor of -174 dBc/Hz at 1 MHz offset frequency, see Fig. 3(a). Although not shown here, the flicker noise could be reduced to -127 $f^{-1}$ dBc/Hz by operating the photodiodes at 10 V and 15 mA of photocurrent. These operating conditions resulted in an increase of the high-frequency noise floor to -170 dBc/Hz at 1 MHz offset, but indicate that reduced power and thermal load on the PD might allow for a lower flicker level. To drive the mixers in the cross-correlator required amplification of one of the 10 GHz signals by 10 dB. Losses from electrical components reduced the net 10 GHz power from each photonic generator by ~ 4 dB. The noise floor beyond 1 MHz is dominated by noise in the 10 dB amplifier, estimated at -178 dBc/Hz, and the thermal noise on the unamplified signal, calculated to be below -178 dBc/Hz. The shot noise contribution to the phase noise of the 10 GHz carrier, reduced due to spectral correlations when detecting ultrashort optical pulses [22], is calculated to be < -200 dBc/Hz.

Figure 3(b) shows the absolute phase noise of two 10 GHz signals obtained via comparison of two independent photonic oscillators. In this measurement, the two MUTC PDs were operated at 21V and 19V (corresponded to an AM-PM null for OFD2), respectively. Each produced 15 mA of photocurrent and +8 dBm microwave power at 10 GHz. As expected, the absolute comparison is noisier than the residual measurement. This is because it reflects noise contributions from the stable optical oscillators and the OFDs. Consequently, for offset frequencies less than 5 kHz, we observe performance similar to our previous measurements [trace (c)] [1]. Amplitude sensitivity of the mixers in the measurement system is believed to have contributed much of the noise from 5 kHz to 1 MHz. The greatest improvement, as compared to Fig. 3(c), was observed at offset frequencies >1 MHz. This is due to the increased signal strength and linearity available with the MUTC PDs. After accounting for the noise contributed by the amplifier, and assuming equal contribution of noise for each OFD, we arrive at a single oscillator phase noise floor of –177 dBc/Hz. Thus low close-in absolute phase noise can be achieved while maintaining a phase noise floor consistent with previous MUTC measurements [22]. The integrated timing jitter of the combined oscillators [trace 3(b)] is only 700 attoseconds (1 Hz to 10 MHz), or about 500 as for a single oscillator, assuming both oscillators are equal. Extending the present noise floor to 5 GHz would increase the single oscillator jitter to 2.6 fs.

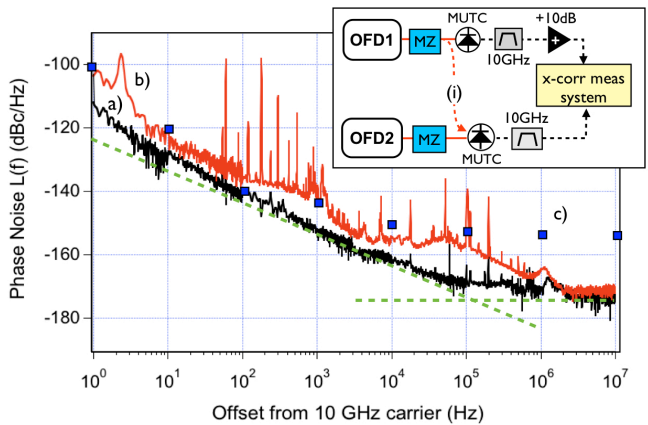

Fig. 3. (Color online) (a) Residual phase noise of two MUTC PDs illuminated by one OFD. (b) Absolute phase noise of two photonic oscillators using MUTC PDs. (c) Results of Ref. [1] for comparison. The dashed green lines indicate the flicker ($f^{-1}$) and high frequency noise floors of trace (a). All traces represent the combined noise of the two systems or PDs. The inset shows the measurement setup. The repetition rate from the OFDs is doubled in a free space Mach-Zehnder (MZ) pulse interleaver. The optical path in the residual measurement is shown by the dashed path (i) with light from OFD2 blocked.

In conclusion, high-power, high-linearity MUTC PDs provide significant advantages for photonic microwave generation in terms of signal strength and low AM-PM sensitivity, demonstrating a 10 GHz microwave phase noise floor nearly 18 orders of magnitude below the carrier. Residual noise measurements of the photodiodes show low flicker noise even when operated at high bias voltage and high photocurrent. Improved thermal management of MUTC PDs [18] and increased optical power should allow for a room temperature photonic generator with a high frequency noise floor competitive with the very best electronic oscillators.

We thank A. Ludlow and T. Rosenband for providing stable CW laser light for the OFDs, and G. Ycas and F. Giorgetta for helpful comments. Financial support is provided by NIST and DARPA. Contribution of the US government. Not subject to copyright in the US.

**References:**
1. T. M. Fortier M. S. Kirchner, F. Quinlan, J. Taylor, J. C. Bergquist, T. Rosenband, N. Lemke, A. Ludlow, Y. Jiang, C. W. Oates, and S. A. Diddams, Nature Photonics **5**, 425 (2011).
2. B. C. Young, F. C. Cruz, W. M. Itano, and J. C. Bergquist, Phys. Rev. Lett. **82**, 3799 (1999).
3. Y. Y. Jiang A. D. Ludlow, N. D. Lemke, R. W. Fox, J. A. Sherman, L. S. Ma, and C. W. Oates, Nature Photonics **5**, 158 (2011).
4. S. A. Diddams A. Bartels, T. M. Ramond, C. W. Oates, S. Bize, E. A. Curtis, J. C. Bergquist, and L. Hollberg, IEEE J. Sel. Top. Quantum Electron. **9**, 1072 (2003).
5. J. J. McFerran, E. N. Ivanov, A. Bartels, G. Wilpers, C. W. Oates, S. A. Diddams, and L. Hollberg, IEEE Electron. Lett. **41** (2005).
6. G. Santarelli P. Laurent, P. Lemonde, A. Clairon, A. G. Mann, S. Chang, A. N. Luiten, and C. Salomon, Phys. Rev. Lett. **82**, 4619 (1999).
7. J. A. Scheer, and J. L. Kurtz, *Coherent Radar Performance Estimation* (Artech House, 1993).
8. G. C. Valley, Opt. Express **15**, 1955 (2007).
9. S. A. Diddams M. Kirchner, T. Fortier, D. Braje, A. M. Weiner, and L. Hollberg, Opt. Express **17**, 3331 (2009).
10. W. Zhang Z. Xu, M. Lours, R. Boudot, Y. Kersale, G. Santarelli, and Y. Le Coq, Appl. Phys. Lett. **96**, 211105 (2010).
11. M. Currie, and I. Vurgaftman, IEEE Photon. Technol. Lett. **18**, 1433 (2006).
12. J. Taylor S. Datta, A. Hati, C. Nelson, F. Quinlan, A. Joshi, and S. Diddams, IEEE Photonics J **3**, 140 (2011).
13. D. A. Tulchinsky, and K. J. Williams, IEEE Photon. Technol. Lett. **17**, 654 (2005).
14. W. Zhang, T. Li, M. Lours, S. Seidelin, G. Santarelli, and Y. Le Coq, Appl. Phys. B **106**, 301 (2012).
15. A. Haboucha W. Zhang, T. Li, M. Lours, A. N. Luiten, Y. Le Coq, and G. Santarelli, Opt. Lett. **36**, 3654 (2011).
16. H. F. Jiang J. Taylor, F. Quinlan, T. Fortier, and S. A. Diddams, IEEE Photonics J **3**, 1004 (2011).
17. Z. Li H. P. Pan, H. Chen, A. Beling, and J. C. Campbell, IEEE J. Quantum Electron. **46**, 626 (2010).
18. Z. Li, Y. Fu, M. Piels, H. P. Pan, A. Beling, J. E. Bowers, and J. C. Campbell, Opt. Express **19**, 385 (2011).
19. T. M. Fortier, A. Bartels, and S. A. Diddams, Opt. Lett. **31**, 1011 (2006).
20. W. F. Walls, in Proceedings of the IEEE Frequency Control Symposium, 257 (1992).
21. T. M. Fortier C. W. Nelson, A. Hati, F. Quinlan, J. Taylor, H. Jiang, C. W. Chou, T. Rosenband, N. Lemke, A. Ludlow, D. Howe, C. W. Oates, and S. A. Diddams, Appl. Phys. Lett. **100** (2012).
22. F. Quinlan T. M. Fortier, H. Jiang, A. Hati, C. Nelson, Y. Fu, J. C. Campbell, and S. A. Diddams, to appear in Nature Photonics.